\documentclass{PoS}

\title{Strange and charm meson masses from twisted mass lattice QCD}

\ShortTitle{Strange and charm meson masses from twisted mass lattice QCD}

\author{\speaker{Martin Kalinowski}\\
         Goethe-Universit\"at Frankfurt am Main, Institut f\"ur Theoretische Physik, \\ 
         Max-von-Laue-Stra{\ss}e 1, D-60438 Frankfurt am Main, Germany \\
         E-mail: \email{kalinowm@th.physik.uni-frankfurt.de}}
\author{Marc Wagner\\
        Goethe-Universit\"at Frankfurt am Main, Institut f\"ur Theoretische Physik, \\ 
        Max-von-Laue-Stra{\ss}e 1, D-60438 Frankfurt am Main, Germany \\
        E-mail: \email{mwagner@th.physik.uni-frankfurt.de}}

\abstract{We present first results of a 2+1+1 flavor twisted mass lattice QCD computation of strange and charm meson masses. We focus on $D$ and $D_s$ mesons with spin $J = 0, 1$ and parity $P = -, +$.}

\FullConference{Xth Quark Confinement and the Hadron Spectrum,\\
		October 8-12, 2012\\
		TUM Campus Garching, Munich, Germany}

\usepackage{pstricks}
\usepackage{pst-plot}
\usepackage{epsfig}
\usepackage{amsmath}

\newcommand{\ltapprox}{\raisebox{-0.5ex}{$\,\stackrel{<}{\scriptstyle\sim}\,$}}

\begin{document}

\section{Lattice setup}

We report about an ongoing project concerned with the computation of the spectrum of mesons with at least one strange or charm valence quark using twisted mass lattice QCD with 2+1+1 dynamical quark flavors. Here we present first results for masses of ground state and excited $D$ and $D_s$ mesons.

Currently we have performed computations on $\approx 400$ gauge field configurations from a single ensemble generated by the European Twisted Mass Collaboration (ETMC). This ensemble is characterized by $24^3 \times 48$ lattice sites, a lattice spacing $a \approx 0.086 \, \textrm{fm}$ and a pion mass $m_\pi \approx 324 \, \textrm{MeV}$. For detailed information regarding these gauge field configurations we refer to \cite{Baron:2010bv}.

For both the valence strange and charm quarks we use degenerate twisted mass doublets, i.e.\ a different discretization as for the corresponding sea quarks. We do this, to avoid mixing of strange and charm quarks, which inevitably takes place in a unitary setup, and which is particularly problematic for hadrons containing charm quarks \cite{Baron:2010th}. The degenerate valence doublets allow two realizations for strange as well as for charm quarks, either with a twisted mass term $+i \mu_{s,c} \gamma_5$ or $-i \mu_{s,c} \gamma_5$. While for a $q \bar{q}$ meson creation operator the sign combinations $(+,-)$ and $(-,+)$ for the quark $q$ and the antiquark $\bar{q}$ are related by symmetry, $(+,+)$ and $(-,-)$, which are also related, yield at finite lattice spacing results slightly different compared to $(+,-)$ and $(-,+)$, due to different discretization errors. Using $(+,-) \equiv (-,+)$ we have tuned the valence strange and charm quark masses to reproduce the physical values of $2m_K^2-m_\pi^2$ and $m_D$, 
two quantities, which only weakly depend on the 
light $u/d$ quark mass.

We have constructed suitable $D$ and $D_s$ meson creation operators using the maximum of sixteen independent $\gamma$ combinations as well as lattice versions of spherical harmonics. Moreover, APE and Gaussian smearing is used, to optimize the overlap of the trial states to the low lying energy eigenstates of interest. More details regarding the construction of meson creation operators in twisted mass lattice QCD can be found in \cite{Jansen:2008si}.

For the computation of the corresponding correlation matrices we use spin diluted timeslice sources in combination with the ``one-end trick''. Meson masses are then determined from plateaux values of corresponding effective masses, which we obtain by ``diagonalizing'' the correlation matrices by solving generalized eigenvalue problems.

\section{Results}

\begin{figure}[t]
	  \centerline{
	  \epsfig{file=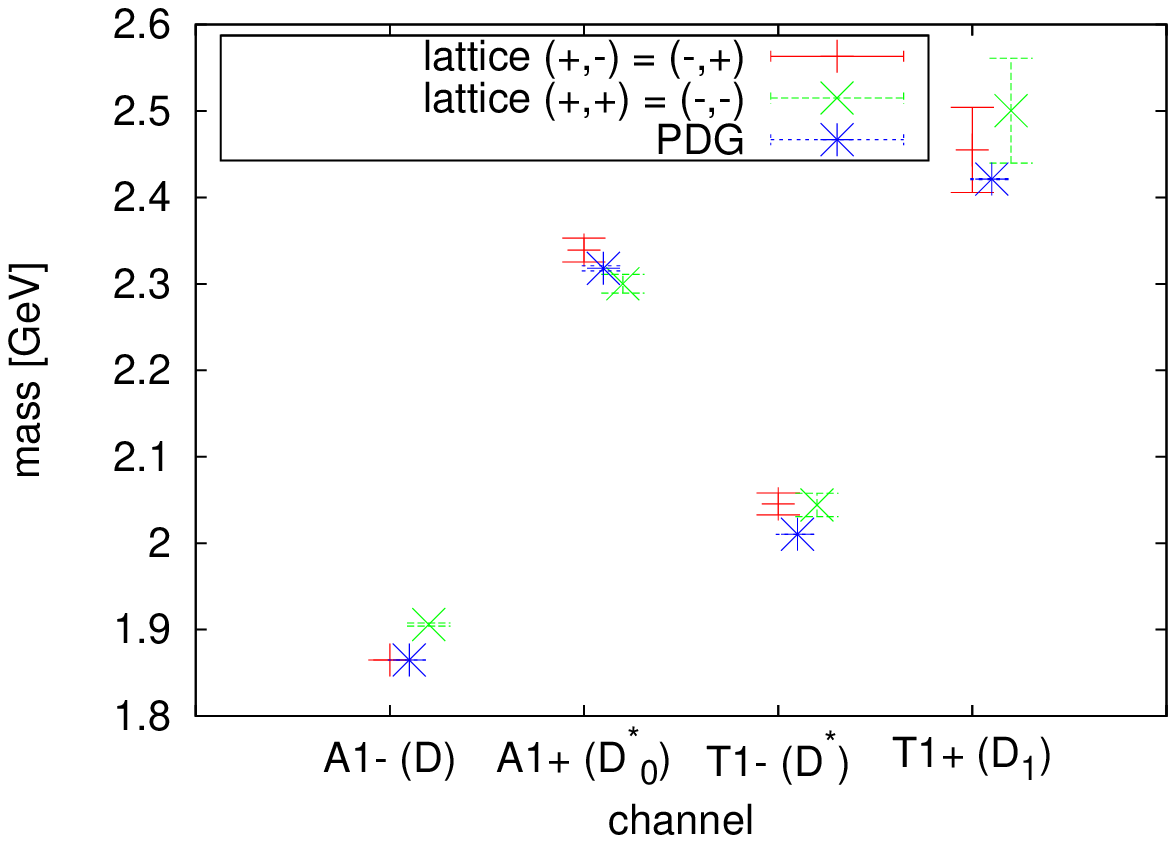, scale=0.61}
          \epsfig{file=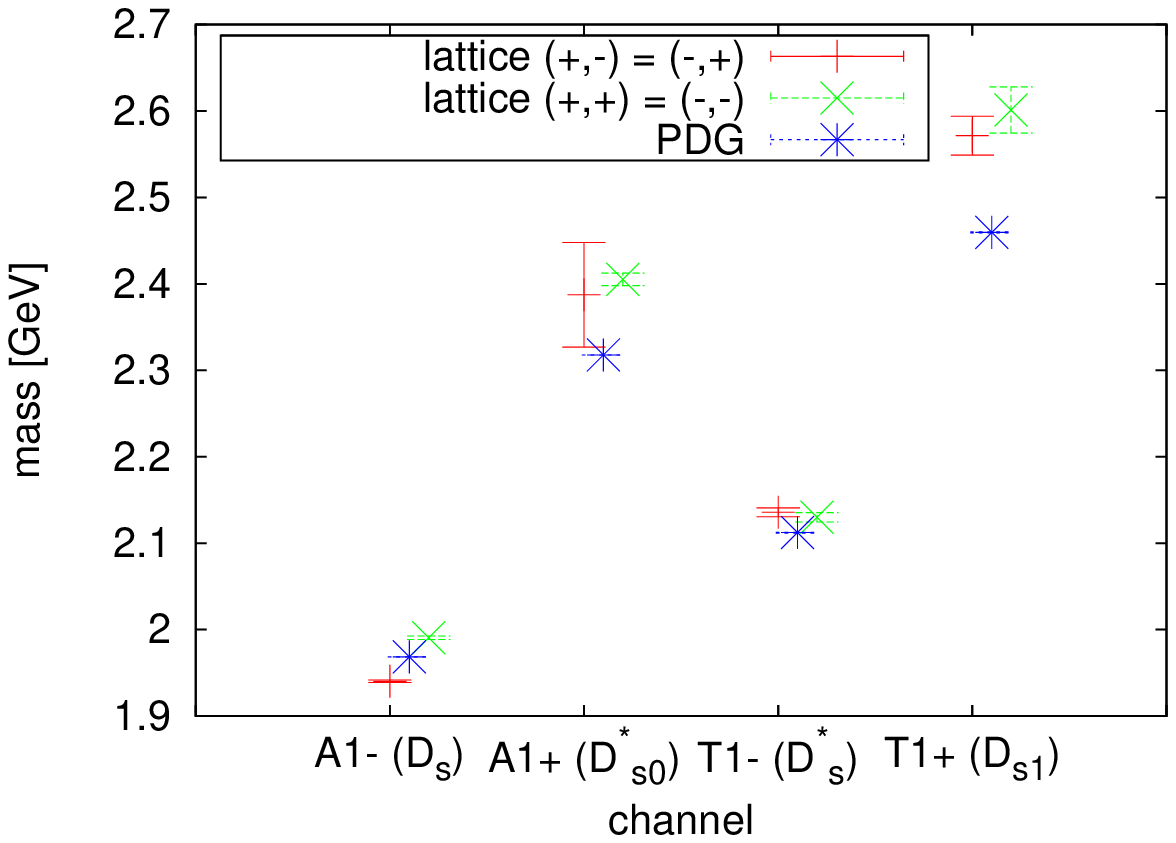, scale=0.61}
	  }
	  \caption{\label{fig:fig1}Low lying $D$ meson (left) and $D_s$ meson (right) spectrum.}
        \end{figure}

In Figure~\ref{fig:fig1} we show $D$ and $D_s$ meson masses for quantum numbers $J^P = 0^-, 0^+, 1^-, 1^+$. Blue data points represent experimental results, while red and green data points correspond to our lattice results obtained with sign combinations $(+,-) \equiv (-,+)$ and $(+,+) \equiv (-,-)$, respectively. The differences between the ``red and green lattice results'' are rather small, indicating that discretization errors at the currently used lattice spacing are $\ltapprox 2\%$.

While in most cases lattice and experimental results agree within statistical errors, we observe significant discrepancies for $D_{s0}^\ast$ and $D_{s1}$. Similar findings have been reported by other lattice studies and by model calculations (cf.\ e.g.\ \cite{Mohler:2011ke,Moir,Ebert:2009ua}). These discrepancies support the assumption that $D_{s0}^\ast$ and $D_{s1}$ are not ordinary quark antiquark states, but might e.g.\ be mesonic $D$-$K$ molecules or diquark-antidiquark pairs (tetraquarks). A corresponding investigation of possibly present four-quark components in positive parity mesons using the same lattice setup is ongoing \cite{Daldrop:2012sr}. A study of the broad $D_0^\ast$ and $D_1$ mesons as resonances using lattice QCD and a composition of usual quark-antiquark states with $D + \pi$  and $D^* + \pi$ can be found in \cite{Mohler:2012na}.

In the near future we plan to extend our computations to states with spin $J > 1$ and to the charmonium spectrum. We will also use ensembles at even smaller lattice spacings to study the continuum limit and we will consider several different pion masses to investigate the light $u/d$ quark mass dependence.

\section*{Acknowledgments}

We thank Christian Wiese for discussions. M.K.\ and M.W.\ acknowledge
support by the Emmy Noether Programme of the DFG (German Research
Foundation), grant WA 3000/1-1 and by Helmholtz Graduate School HGS-HIRe for FAIR.
This work was supported in part by the Helmholtz International Center
for FAIR within the framework of the LOEWE program launched by the State
of Hesse.

\end{document}